\def\cm{\rm cm$^{-1}$}
\def\pf6{(TM\-TSF)$_2$\-PF$_6$}
\def\asf6{(TM\-TSF)$_2\-$AsF$_6$}
\def\clo4{(TM\-TSF)$_2$\-ClO$_4$}

\documentclass[global]{ltxguide}[1995/11/28]
\usepackage{epsfig}

\title{Ordering phenomena
in  quasi one-dimensional organic conductors}

\author{Martin Dressel\\
1.\ Physikalisches Institut, Universit\"at Stuttgart,\\
Pfaffenwaldring 57, D-70550 Stuttgart, Germany\\
Phone: +49-711-685 64946, FAX: +49-711-685 64886,\\ email:~dressel@pi1.physik.uni-stuttgart.de}

\date{\today}

\begin{document}

\maketitle
\begin{abstract}
Low-dimensional organic conductors could establish themselves as model systems for the  investigation of the physics in reduced dimensions. In the metallic state of a one-dimensional solid, Fermi-liquid theory breaks down and spin and charge degrees of freedom become separated. But the metallic phase is not stable in one dimension: as the temperature is reduced, the electronic charge and spin tend to arrange themselves in an ordered fashion due to strong correlations. The competition of the different interactions is responsible for which broken-symmetry ground state is eventually realized in a specific compound and
which drives the system towards an insulating state.
Here we review the various ordering phenomena and how they can be identified by optic and magnetic measurements. While the final results
might look very similar in the case of a charge density wave and a charge-ordered metal, for instance, the physical cause is completely different. When density waves form, a gap opens in the density of states at the Fermi energy due to nesting of the one-dimension Fermi surface sheets.
When a one-dimensional metal becomes a charge-ordered Mott insulator, on the other hand, the
short-range Coulomb repulsion localizes the charge on the lattice sites and even causes certain charge patterns.
We try to point out the similarities and conceptional differences of these phenomena and give an example for each of them.
Particular emphasis will be put on collective phenomena which are inherently present as soon as ordering breaks the symmetry of the system.
\end{abstract}


\section{Introduction}
Physics in one dimension is a fascinating topic for theory and experiment.
One-dimensional models are simpler compared to three-dimensional ones and in many cases can be solved analytically only then (Lieb and Mattis 1966). Often the reduction of dimensionality does not really matter because the essential physics remains unaffected. But there are also a number of phenomena in condensed matter which only or mostly occur in one dimension. %
In general, the dominance of the lattice is reduced and electronic interactions become superior. Quantum mechanical effects are essential as soon as the confinement approaches the electron wavelength. Fundamental concepts of physics, like the Fermi liquid theory of interacting particles breaks down in one dimension and has to be replaced by alternative concepts based on collective excitations (Giamarchi 2004).

One-dimensional structures are intrinsically unstable for thermodynamic reasons. Hence various kinds of ordering phenomena may take place which break the translational symmetry of the lattice, charge or spin degrees of freedom: Phase transitions occur as a function of temperature or some order parameter.
On the other hand, fluctuations suppress long-range order at any finite temperature in one (and two) dimension. The ordered ground state is only stabilized by the fact that real systems consist of one-dimensional chains, which are coupled to some degree. The challenge now is to extract the one-dimensional physics from experimental investigations of quasi-one-dimensional systems and to check the theoretical predictions.
Besides pure scientific interest, the crucial importance of these phenomena in nanotechnology might not lie  ahead too far.

After a short overview of how quasi-one-dimensional structures can be achieved in reality, model systems of organic conductors are presented.
The different ordering phenomena are introduced by simple and intuitive pictures and elucidated by typical examples. As far as the charge degree of freedom is concerned, charge density waves and charge order break the translational invariance and drive a metal-insulator-transition.
Magnetic or non-magnetic ground states are obtained when the electron spins order in a certain manner. In any case, the physical properties
change dramatically upon the phase transition and new collective phenomena occur.
This review tries not to get lost in the richness of the observations by pointing out the common concepts.

\section{Realization of one-di\-men\-sion\-al structures}
There are numerous ways to approximate one-dimensional physics in reality.
The ideal one-dimensional system would be an infinite chain of atoms in vacuum; close enough to interact with their neighbors, but completely isolated from the environment. Over the past years significant progress has been made towards the realization of one-dimensional atomic gases, based on Bose-Einstein condensates of alkalides trapped in two-dimensional optical lattices (Moritz et al. 2003). Unfortunately this technique is far from being readily available as a versatile tool for broad investigations. Hence the most obvious approach to achieve one-dimensional physics in solids would be to utilize advanced semiconductor technology (Davies 1998). Besides the enormous technological effort, this approach has the disadvantage that these structures are embedded in bulk materials and not easily accessible to further experiments.

If the surface of a single crystal, like silicon, is cut in a small angle with respect to a crystallographic direction, terraces are produced with mono-atomic steps separating them. The surface reconstruction may lead to an anisotropic arrangement with the possibility of one-dimensional structures. Evaporating gold on top of it, the atoms organize themselves in rows along these steps, forming atomic wires (Himpsel et al. 2001).

It is possible to grow bulk materials as extremely thin and long hair-like crystals when stress is applied; they are known as whiskers of gold, silver, zinc, tin, etc. While metallic whiskers often lead to circuit shortages and failures and sought to be avoided,
enormous potential of applications is seen in another sort of filaments: carbon nanotubes. This fascinating and versatile material, which was discovered about ten years ago,
solely consists of carbon atoms. Nanotubes can be considered as rolled-up sheets of graphite with electrical properties very much depending on the winding ratio. Single-wall carbon nanotubes with a small diameter and the right winding ratio are excellent realizations of one-dimensional conductors (O'Connell 2006).

By far the most successful approach to one-di\-men\-sion\-al physics are highly anisotropic crystals. Here the Krogmann salt K$_2$Pt(CN)$_4$Br$_{0.3}\cdot$H$_2$O, known as KCP, probably represents the most intuitive example for it consists of a simple chain of platinum ions with overlapping $d$ orbitals (Zeller 1973, 1975). Alternatively, transition metal oxides are known for decades to crystallize in low-dimensional structures (Monceau 1985).
Varying the composition and structural arrange\-ment provides the possibility to obtain one- and two-dimensional conductors or superconductors, but also spin chains and ladders (Vuleti\'c et al. 2006). The interplay of the different degrees of freedom together with the importance of electronic correlations makes these systems an almost unlimited source for novel and exciting phenomena and a challenge for their theoretical understanding (Maekawa et al. 2004).

\section{Organic conductors}
While in KCP the metallic properties are due to the platinum ions, organic conductors constitute a class of solids with no metal atoms present (or relevant); instead the $\pi$ electrons distributed over of the entire organic molecule form the orbitals, which might overlap and lead to band-like conductivity. The additional degree of freedom, tailoring these molecules, supplements the structural arrangement in the crystal and makes it possible to fine-tune competing interactions from the various degrees of freedom. This makes organic materials superior for studying low-dimensional physics and ordering phenomena in solids (Farges 1994; Ishiguro, Yamaji and Saito 1998, Batail 2004).

\begin{figure}
\centerline{\epsfig{file=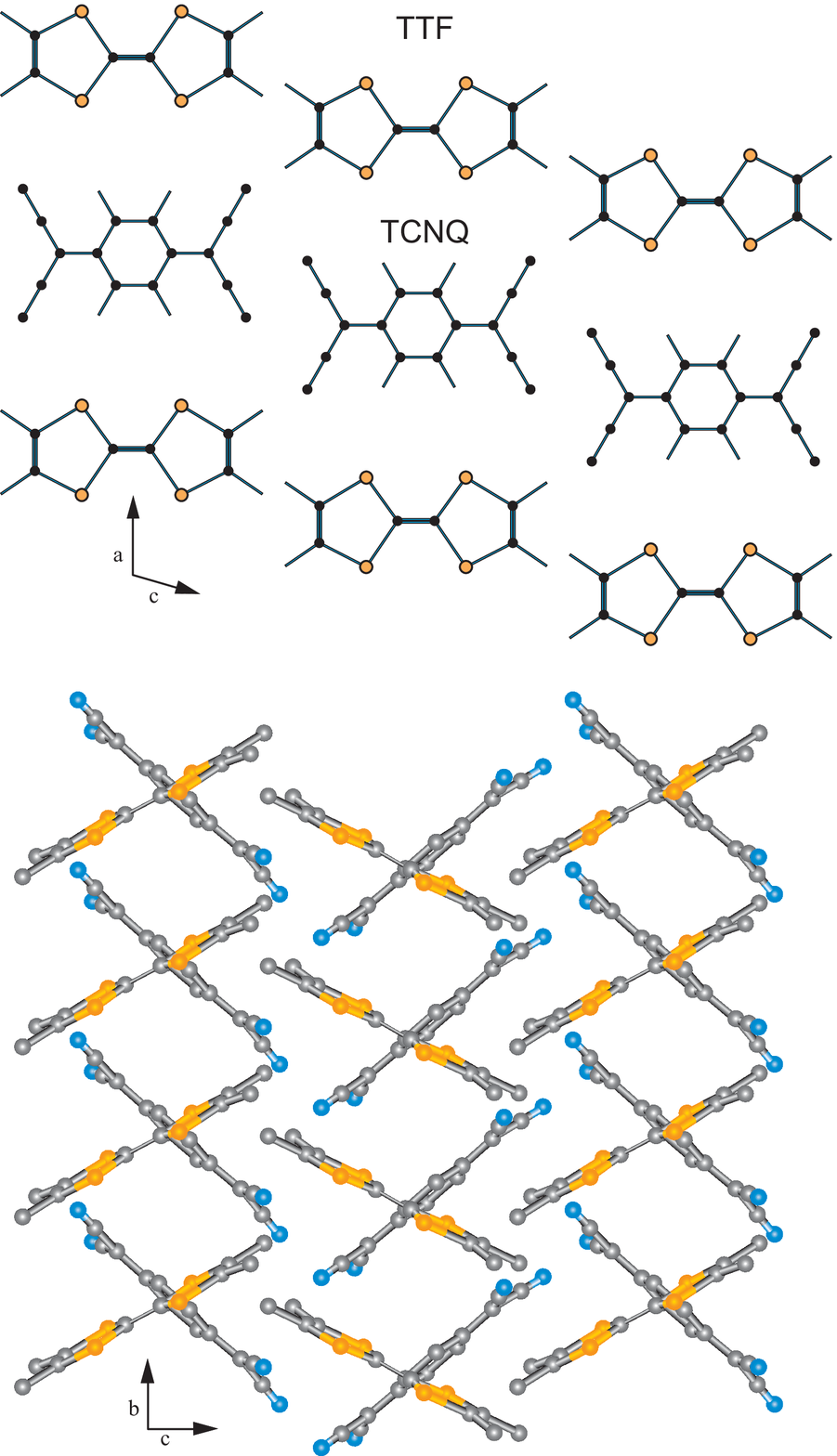,width=6cm}}
 \caption{\label{fig:structureTTFTCNQ}
\small (a)~The TTF molecules and the TNCQ molecules form separate stacks
along the $b$-direction. (b)~The molecules are tilt in herring bone fashion. The $\pi$ orbitals overlap in the $b$-direction and form the conduction band that causes the metallic properties along the stacks.}
\end{figure}

In general these synthetic metals consist of piles of planar molecules with the atomic orbitals overlapping along the stack.
In the perpendicular directions the conductivity is orders of magnitude lower
because the distance between the stacks is large and in addition they may be separated by couterions.
There are two prerequisite for a good electronic transport:
the overlap of the orbitals and an electronic charge transfer between donor and acceptor molecules to generate partially filled bands.
The breakthrough of organic conductors happened in the early 1970s with the synthesis of
\underline{t}etra\-\underline{t}hio\-\underline{f}ulvalene-\underline{t}etra\-\underline{c}ya\underline{n}o\-\underline{q}uino\-methane
which exhibits a room temperature conductivity of $10^3~(\Omega{\rm cm})^{-1}$ and an anisotropy of more than a factor of 100
(Coleman et al. 1973, Cohen et al. 1974).
TTF-TCNQ is a charge-transfer compound with separate stacks of
the cations TTF (charge donors) and anions TCNQ (electron acceptors), as depicted in Fig.~\ref{fig:structureTTFTCNQ}. It has very good metallic properties down to a temperature of approximately 60~K where a metal-insulator transition occurs .

\begin{figure}
\centerline{\epsfig{file=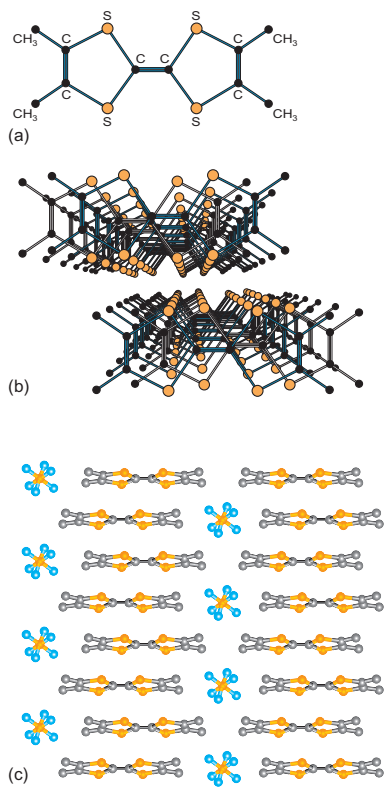,width=5cm}}
 \caption{\label{fig:structureTMTTF}
\small (a) TMTTF molecule
(b) View along the stacks of TMTTF
($a$-direction) and (c) perpendicular to them ($b$-direction).
Along the $c$-direction the stacks of the
organic molecules are separated by the AsF$_6^-$ anions, for instance.
In the case of the TMTSF salts, S is replaced by Se.}
\end{figure}

In the course of the last two decades, in particular the Bechgaard salts \underline{t}etra\-\underline{m}ethyl-\underline{t}etra\-\underline{s}elena\-\underline{f}ulvalene
(TMTSF), and its variant TMTTF where selenium is replaced by sulfur, turned out to be an excellent model for quasi-one-dimensional metals, superconductors, charge order, spin-density-wave systems, spin chains, spin-Peierls systems, etc.\ depending on the degree of coupling along and perpendicular to the chains (J{\'e}rome and Schulz 1982, Dressel 2003).
The  planar organic molecules stack along the $a$-direction with
a distance of approximately 3.6~\AA.
In the $b$-direction the coupling between the
chains is relatively small (but not negligible), and in the third direction the stacks are even separated by the inorganic anion, like
PF$_6^-$, AsF$_6^-$, ClO$_4^-$, Br$^-$, etc. as depicted  in Fig.~\ref{fig:structureTMTTF}.
Each organic molecule transfers half an electron to the counterions yielding a quarter-filled hole band.
In general a small dimerization creates pairs of organic molecules; the conduction band gets split. In addition, spontaneous charge disproportionation, called charge ordering (CO)
may divide the molecules into two non-equivalent species (cf. Fig.~\ref{fig:coso}) commonly observed in TMTTF salts.
Due to the instability of the quasi one-dimensional Fermi surface, at ambient pressure
(TMTSF)$_2$PF$_6$ undergoes a transition to a spin-density-wave (SDW) ground state at $T_{\rm SDW} = 12$~K.
Applying pressure or
replacing the PF$_6^-$ anions by ClO$_4^-$ leads to a stronger coupling in the second direction: the
material becomes more two-dimensional. This seems to be a requirement for superconductivity (J{\`e}rome et al. 1980, J{\'e}rome and Schulz 1982).

In an enormous effort by research groups all around the world,
the family of TMT$C$F salts (where $C$ is one of the chalcogenes
selenium or sulfur)
was intensively explored and became the model system of quasi one-dimensional
conductors. By external pressure or substitution of anions (chemical pressure)
the interchain coupling increases and thus the dimensionality
crosses over from a strictly one-dimension\-al to a more two or
three-dimensional system. Over the last two decades various groups contributed to
the rich phase diagram as displayed in Fig.~\ref{fig:phasediagram}.
Besides the Mott insulating state,  spin
Peierls,  antiferromagnetic insulator, spin-density-wave,
and superconductivity, also the metallic state changes
its behavior going from a Luttinger liquid (in strictly one dimension) to a Fermi liquid (in higher dimensions);
these properties were summarized in a recent review (Dressel 2003).

\section{Charge degree of freedom}
At first glance, there seems to be no good reason that in a chain of molecules the sites are not equivalent, or that the itinerant charges of a one-dimensional metal are not homogeneously distributed. However, the translational symmetry can be broken if electron-phonon interaction and electron-electron interaction become strong enough; later we will also consider spin-phonon coupling. Energy considerations then cause a charge redistribution in one or the other way, leading to charge density waves or charge order.
Indeed, these ordering phenomena affect most thermodynamic, transport and elastic properties of the crystal, and in some cases also its structure; here we want to focus on the electrodynamic response, i.e.\ optical properties in a broad sense.

Any sort of charge disproportionation implies a partial localization of the electrons. The density of states at the Fermi level is reduced which has severe consequences for the metallic state. In certain cases the material can even become totally insulating with a complete gap open.
First of all, there will be single-particle electron-hole excitations which require an energy of typically an eV, like in a band insulator. But in addition, collective modes are expected. There is a rather general argument by Goldstone (1961) that whenever a continuous symmetry is broken, long-wavelength modulations in the symmetry direction should occur at low frequencies.
The fact that the lowest energy state has a broken symmetry means that the system is stiff: modulating the order parameter (in amplitude or phase) will cost energy. In crystals, the broken translational order introduces a rigidity to shear deformations, and low-frequency phonons. These collective excitations are expected well below a meV.

\subsection{Charge density wave}
\begin{figure}
\centerline{\epsfig{file=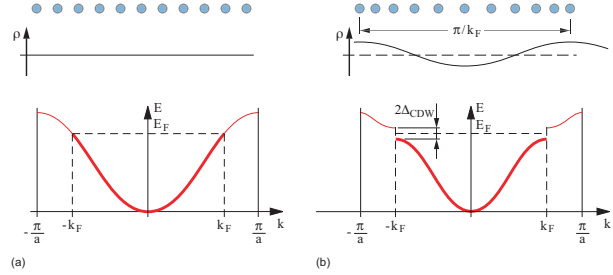,width=8cm}}
\caption{\label{fig:cdw1}
\small (a)~In a regular metal, the charge is homogeneously distributed in space. The conduction band is filled up to the Fermi energy $E_F$. (b)~A modulation of the charge density with a wavelength $\lambda=\pi/k_F$ changes the periodicity; hence in $k$-space the Brillouin zone is
reduced which causes a gap $2\Delta_{\rm CDW}$ at $\pm k_F$. The system becomes insulating.}
\end{figure}

The energy dispersion forms electronic bands which are filled up to the Fermi wave-vector $\bf k_F$. In one dimension, the Fermi surface consists of only two sheets at $\pm k_F$. The crucial point is that the entire Fermi surface can be mapped onto itself by a $2k_F$ translation.
Since the density of states in one dimension diverges as $(E-E_0)^{-1/2}$ at the band-edge $E_0$, the electronic system is very susceptible to $2k_F$ excitations. The result of the Fermi surface nesting and divergency of the electronic density of states is a spatial modulation in the charge density $\rho({\bf r})$ with a period of $\lambda=\pi/k_F$ (Fig.~\ref{fig:cdw1}), which does not have to be commensurate to the lattice: this is called a charge density wave (CDW).
 Long-range charge modulation is crucial because a CDW is a $k$-space phenomenon. Mediated by electron-phonon coupling, this causes a displacement of the underlying lattice (Peierls instability). The gain in electronic energy due to the lowering of the occupied states has to over-compensate the energy required to modulate the lattice (Monceau 1985, Gr\"uner 1994).

The consequence of the CDW formation is an energy gap $2\Delta_{\rm CDW}$ in the single particle excitation spectrum, as observed in the activated behavior of electronic transport or a sharp onset of optical absorption.
Additionally, collective excitations are possible which allow for translation of the density wave as a whole. Although pinning to lattice imperfections prevents Fr\"ohlich superconductivity, the density-wave ground state exhibits several spectacular features, like a pronounced non-linearity in the charge transport (sliding CDW) and a strong oscillatory mode in the GHz range of frequency (pinned-mode resonance) (Gr\"uner 1994).

\begin{figure}
\centerline{\epsfig{file=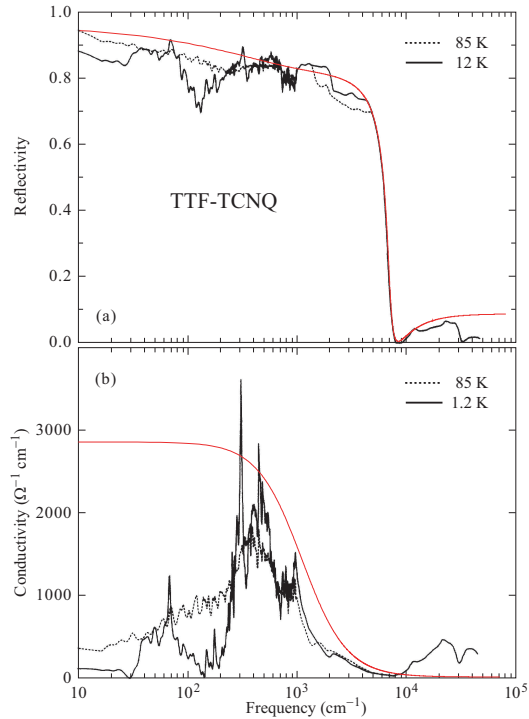,width=7cm}}
\caption{\label{fig:cdw2}
\small (a)~Optical reflectivity and (b)~conductivity of TTF-TCNQ parallel to the stack direction $b$ for temperatures above and below the charge density wave transition $T_{\rm CDW} = 53$~K (data taken from Basista et al. 1990). The insulating state is seen by the drop in the low-frequency reflectivity. The suppression of the conductivity below 300~\cm\ indicates the opening of the CDW gap. For comparison, the thin red lines represent the simple metallic behavior according to the Drude model with a plasma frequency of $\omega_p/(2\pi c)=42\,000~{\rm cm}^{-1}$ and a scattering rate of $1/(2\pi\tau c)=1200~{\rm cm}^{-1}$.}
\end{figure}

The theory of CDW was triggered by the observations made on the charge-transfer salt TTF-TCNQ in the 1970s and 80s (Denoyer et al. 1975, Heeger and Garito 1975, Kagoshima et al. 1988), in spite of the fact that the compound exhibits a more complex scenario in the temperature range between 60~K and 33~K due to fluctuations, separate ordering in the different stacks, transverse coupling, commensurate and incommensurate phases etc.
In Fig.~\ref{fig:cdw2} the optical properties of TTF-TCNQ are presented as an example (Basista et al. 1990). Clear deviations from the Drude behavior of a conventional metal (Dressel and Gr{\"u}ner 2002) are observed due to the one-dimensional nature; a similar behavior has been reported for the Bechgaard salts (TMTSF)$_2X$ (Dressel 2003). When the temperature is reduced below the $T_{\rm CDW}\approx 53$~K, the low-frequency reflectivity drops because an energy gap opens at the Fermi level. The single-particle gap opens around 290~\cm, but in addition a strong mode is found between 50 and 100~\cm. The explanation of this feature as well as indications of the pinned mode resonance at even lower frequencies is still under debate.


\subsection{Charge order}
\begin{figure}
\centerline{\epsfig{file=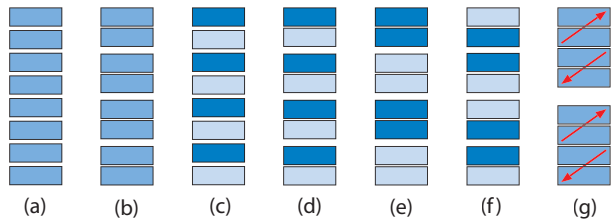,width=8cm}}
\caption{\label{fig:coso}
\small (a)~Stack of equally spaced organic molecules with charge $\rho=\rho_0=1/2 e$ per molecule, for instance;
(b)~the dimerization of the stack leads to alternating distances between adjacent molecules. (c)~Charge ordering modulates the
electronic charge $\rho$ per molecule to $\rho_0+\delta$ and $\rho_0 -\delta$;
as indicated by the gray values. (d)~The charge disproportionation can be accompanied by a lattice dimerization;
besides a ({--} +   {--} +) pattern, also the ({--} {--}  + +) alternation is possible in two different fashions: (e)~({--} {--} $\vert$ + + $\vert$ {--} {--} $\vert$ + +)  and (f)~( + {--}  $\vert$ {--} +  $\vert$ + {--}  $\vert$ {--} +). (g)~In addition the magnetic moments located at of the molecules can pair and form spin singlets.}
\end{figure}

The crucial point of a CDW is the Fermi surface nesting; the driving force is the energy reduction of the occupied states right below the Fermi energy $E_F$ when the superstructure is formed (cf. Fig.~\ref{fig:cdw1}). Well distinct from a charge density wave is the occurrence of charge order (CO).
The Coulomb repulsion $V$ between adjacent lattice sites may lead to the preference of alternatingly more or less charge as depicted in Fig.~\ref{fig:coso}c. The extended Hubbard model is a good description of the relevant energies (Seo and Fukuyama 1997):
\begin{eqnarray}
{\cal H}= &-&t\sum_{j=1}\sum_{\sigma=\uparrow\downarrow} \left(c^+_{j,\sigma}c_{j+1,\sigma}
+ c^+_{j+1,\sigma}c_{j,\sigma}\right)\nonumber\\
&+& U\sum_{j=1} n_{j\uparrow}n_{j\downarrow} +
V \sum_{j=1}  n_{j} n_{j+ 1} \label{eq:Hubbard} \quad .
\end{eqnarray}
Here $t$ denotes the hopping integral to describe the kinetic energy, $U$ is the on-site Coulomb repulsion and $V$ the nearest neighbor interaction.
The disproportionation of charge on the molecules represents a short-range order and has to be commensurate with the lattice. CO may be accompanied by a slight lattice distortion (Fig.~\ref{fig:coso}d), but this is a secondary effect.
In contrast to a CDW, a metallic state above the ordering temperature is not required. If it is the case (metallic state), the gap in the density of states due to the superstructure also causes a metal-insulator transition.

\begin{figure}
\centerline{\epsfig{file=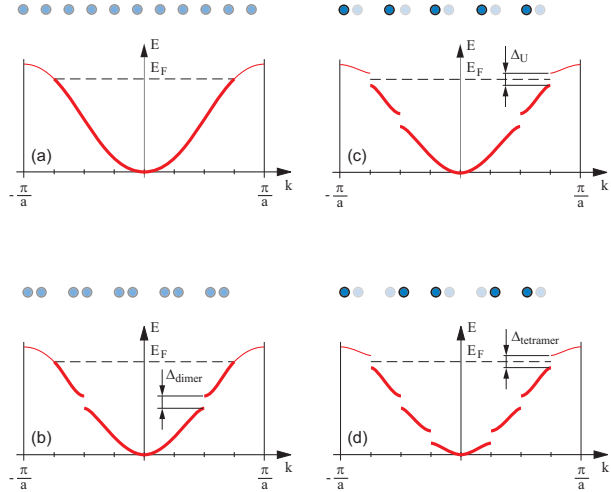,width=8cm}}
\caption{\label{fig:co1}
\small (a)~A homogeneous stack of TMT$C$F, for example, with half an electronic charge $+e$  per molecule results in a three-quarter-filled band which leads to metallic behavior. (b)~Dimerization doubles the unit cell and the Brillouin zone is cut into two equal parts. The upper band is half filled
and the physical properties remain basically unchanged. (c)~Due to Coulomb respulsion $U$ a gap $\Delta_{\rm U}$ opens at the Fermi energy $E_F$ that drives a metal-to-insulator transition. (d)~The tetramerization doubles the unit cell again.}
\end{figure}

The quasi-one-dimensional (TMTTF)$_2X$ salts are poor conductors at ambient temperature and exhibit a rapidly increasing resistivity as the temperature is lowered (Fig.~\ref{fig:dc}). The reason is the accumulation of two effects which severely influence the energy bands as depicted in Fig.~\ref{fig:co1}. The first one is a structural: due to the interaction with the anions (Fig.~\ref{fig:structureTMTTF}) the molecular stack is dimerized as visualized in Fig.~\ref{fig:coso}b. The conduction band is split by a dimerization gap $\Delta_{\rm dimer}$ and the material has a half-filled band. In a second step the Coulomb repulsion $V$ causes charge disproportionation within the dimers (Fig.~\ref{fig:coso}d). This also drives the one-dimensional half-filled system towards an insulating state: correlations induce a gap $\Delta_{\rm U}$ at the Fermi energy $E_F$ as shown in Fig.~\ref{fig:co1}c. The tetramerization of the CO according to Fig.~\ref{fig:coso}e and f changes this picture conceptually (Fig.~\ref{fig:co1}d): the soft gap $\Delta_{\rm CO}$ due to short-range nearest-neighbor interaction $V$ localizes the charge carriers. If not completely developed it just results in a reduction of the density of state (pseudogap). The tetramerization gap, on the other hand, is related to long-range order.

One- and two-dimensional NMR spectroscopy demonstrated the existence of an intermediate charge-ordered phase in the TMTTF family. At ambient temperature, the spectra are characteristic of nuclei in equivalent molecules. Below a continuous charge-ordering transition temperature $T_{\rm CO}$, there is evidence for two inequivalent molecules with unequal electron densities. The absence of an associated magnetic anomaly indicates only the charge degrees of freedom are involved and the lack of evidence for a structural anomaly suggests that charge-lattice coupling is too weak to drive the transition (Chow et al. 2000).

The first indications of CO came from dielectric measurements in the
radio-frequency range (Nad et al. 1999, Monceau et al. 2001) where a divergency of the low-frequency
dielectric constant was observed at a certain temperature $T_{\rm CO}$.
This behavior is well known from ferroelectric transitions, where also
a dielectric catastrophe is observed due to a softening of the lattice.
The idea is that at elevated temperatures the molecules carry
equivalent charge of $+0.5e$; but upon lowering the temperature, the
charge alternates by $\pm\rho$ causing a permanent dipole moment. On
this ground new intermolecular vibrations at far-infrared frequencies
below 100~cm$^{-1}$ become infrared active along all three crystal axes
in the CO state due to the unequal charge distribution on the TMTTF
molecules. Above the CO transition these modes, which can be assigned
to translational vibrations of the TMTTF molecules, are infrared silent
but Raman active. In (TMTTF)$_2$AsF$_6$, for instance, we observe a
strong vibration around 85~\cm\ for $E\parallel a$, at 53~\cm\ and
66~\cm\ for the $b$ and $c$ directions, respectively, as soon as
$T<100$~K. By now there are no reports on a
collective excitation which should show up as a low-frequency phonon.

\begin{figure}
\centerline{\epsfig{file=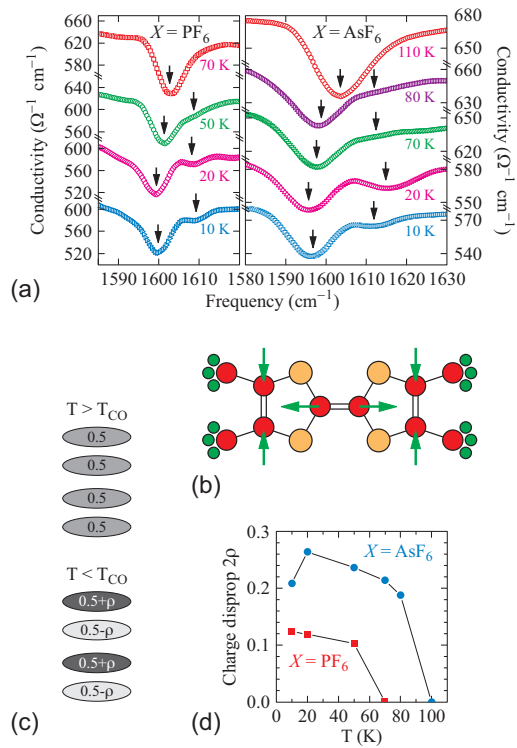,width=7cm}}
\caption{\label{fig:co3}
\small (a)~Mid-infrared conductivity of (TMTTF)$_2$\-PF$_6$ and (TMTTF)$_2$\-AsF$_6$ for light polarized parallel to the molecular stacks. (b)~The emv coupled totally-symmetric intramolecular $\nu_3$(A$_g$) mode (which mainly involves the C=C double bonds) is
very sensitive to the charge carried by the molecule; hence
the mode splits due to charge order. (c)~While the molecules contain equal charge for $T>T_{\rm CO}$, an alternation of charge-rich and charge-poor sites is found as the crystal is cooled below $T_{\rm CO}$. (d) The disproportionation in charge increases with decreasing temperature  (after Dumm et al. 2005).}
\end{figure}

The CO can be locally probed by intramolecular vibrations. Totally symmetric ${\rm A}_g$ modes are not infrared active; nevertheless due to electron-molecular vibrational (emv) coupling (i.e.\ the charge transfer between two neighboring organic TMTTF molecules which vibrate out-of phase) these modes can be observed by infrared spectroscopy for the polarization parallel to the stacks ($E\parallel a$). As demonstrated in Fig.~\ref{fig:co3}, the resonance frequency is a very sensitive measure of the charge per molecule (Dumm et al. 2005). The charge disproportionation increases as the temperature drops below $T_{\rm CO}$ in a mean-field fashion expected from a secon-order transition; the ratio amounts to about 2:1 in (TMTTF)$_2$AsF$_6$ and 5:4 (TMTTF)$_2$PF$_6$. The charge disproportionation is slightly reduced in the AsF$_6$ salt, when it enters the spin-Peierls state, and unchanged in the antiferromagnetic PF$_6$ salt which infers the coexistence of charge order and spin-Peierls order at low temperatures.

\subsection{Neutral-ionic transition}
\begin{figure}
\centerline{\epsfig{file=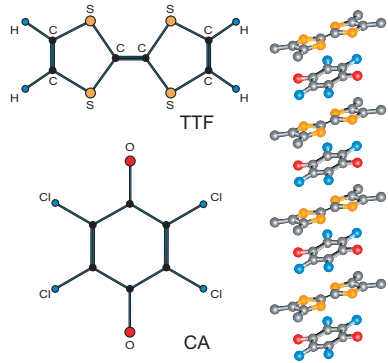,width=5cm}}
 \caption{\label{fig:structureTTFCA}
\small The TTF and chloranil QCl$_4$ are planar molecules. In the mixed-stack compound TTF-CA the two distinct molecules  alternate.}
\end{figure}

While in the previous example the crystals consist of separate cation
and anion chains between which the electron transfer occurs,
mixed-stack organic charge-transfer compounds have only one type of
chain composed of alternating $\pi$ electron donor and acceptor
molecules (... A$^{-\rho}$D$^{+\rho}$A$^{-\rho}$
D$^{+\rho}$A$^{-\rho}$D$^{+\rho}$ ...) as sketched in Fig.~\ref{fig:structureTTFCA}. These materials are either
neutral or ionic, but under the influence of pressure or temperature
certain neutral compounds become ionic. There is a competition between
the energy required for the formation of a D$^+$A$^-$ pair and the
Madelung energy. Neutral-ionic (NI) phase transitions are collective,
one-dimensional charge-transfer phenomena occurring in mixed-stack
charge-transfer crystals, and they are associated to
many intriguing phenomena, as the dramatic increase in conductivity and
dielectric constant at the transition (Torrance et al. 1981, Horiuchi et al. 2000).

In the simplest case, the charge per molecule changes from completely neutral $\rho=0$ to fully ionized $\rho=1$. Ideally this redistribution of charge is decoupled from the lattice, and therefore should not change the inter-molecular spacing. In most real cases, however, the
NI transition is characterized by the complex interplay between the average ionicity $\rho$ on the molecular sites and the stack
dimerization $\delta$.
The ionicity may act as an order parameter only
in the case of discontinuous, first order phase transitions.
While the inter-site Coulomb interaction $V$ favors a discontinuous jump of ionicity, the intra-chain charge-transfer integral $t$ mixes
the fully neutral and fully ionic quantum states and favors continuous changes in $\rho$.
The coupling of $t$ to lattice phonons induces the dimerization of the
stack, basically a Peierls-like transition to a ferroelectric state,
which is a second order phase transition.
Intramolecular (Holstein) phonons, on the other hand, modulate the
on-site energy $U$ and favor a discontinuous jump in $\rho$.

\begin{figure}
\centerline{\epsfig{file=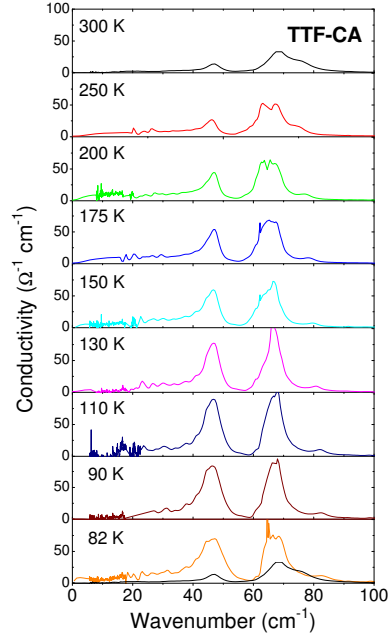,width=5cm}}
\caption{\label{fig:TTF-CAcond}
\small Low-frequency conductivity of
TTF-CA for $T>T_{\rm NI}$ for different temperatures as indicated in the
panels. As the NI transition is approached by decreasing temperature, the modes become stronger and an additional band appears as low as 20~cm$^{-1}$.
To make the comparision easier, the room temperature spectrum (black line) is replotted in the lowest frame.}
\end{figure}

The temperature induced NI transition of tetrathiafulvalene-tetrachloro-{\it p}-benzo\-quinone
(TTF-CA) at $T_{\rm NI}=81$~K is the prime example of a first-order
transition with a discontinuous jump in $\rho$. This can
be seen be an abrupt change in the optical properties;
below the NI transition the coupled bands
shift to higher frequencies (Masino et al.\ 2006). In terms of a modified, one-dimensional Hubbard model [similar to Eq.~(\ref{eq:Hubbard})], the NI transition can
be viewed as a transition from a band insulator to a Mott insulator due to the competition between the energy difference between donor and acceptor
sites, and the on-site Coulomb repulsion  $U$. Peierls and Holstein
phonons are both coupled to charge transfer electrons, albeit before
the NI transition the former are only infrared active, and the latter
only Raman active. This makes polarized Raman and reflection
measurements a suitable tool to explore the NI transition. The optical
experiments identify practically all the totally symmetric modes of
both neutral and ionic phases of TTF-CA. The vibronic bands present in
the infrared spectra for $T>T_{\rm NI}$ are due to sum and
difference combinations involving the lattice mode, which gives rise to
the Peierls distortion at the transition. In Fig.~\ref{fig:TTF-CAcond}
the low-frequency conductivity spectra are plotted for different
temperatures $T>T_{\rm NI}$. From calculations we expect three lattice
modes which couple to electrons and become stronger as the transition
is approached. The lattice modes strongly couple to electrons and
behave as soft modes of the ferroelectric transition at $T_{\rm
NI}=81$~K. The lowest mode softens most and is seen strongly overdamped
around 20~cm$^{-1}$. The temperature evolution of this Peierls mode,
which shows a clear softening (from 70 to 20~cm$^{-1}$) before the
first-order transition to the ionic ferroelectric state takes place.
In the ordered phase a clear identification and theoretical modelling of the Goldstone mode is still an open problem because the system has several degrees of freedom coupled to each other.

The cooperative charge transfer among the constructive molecules of TTF-CA can also be induced by irradiation of a short laser pulse. A photoinduced local charge-transfer excitation triggers the phase change and cause the transition in both directions (Koshihara et al. 1999). When Cl is replaced by Br in the tetrahalo-{\it p}-benzoquinones the lattice is expanded, (like a negative pressure) and the ionic phase vanishes completely. Hydrostatic pressure or Br-Cl substitution is utilized as a control parameter to more or less continuously tune the NI transition at $T\rightarrow 0$ (Horiuchi et al. 2003).

\section{Spin degree of freedom}
In addition to the charge, electrons carry also a spin, which can interact with each other and with the underlying lattice. It was shown by Overhauser in the early 60s that an electron gas is instable and forms a spin density wave, i.e.\ the spin of the itinerant electrons order antiferromagnetically. But also localized magnetic moments can undergo some ordering due to electronic interaction as known from transition metal compounds, for instance. The ground state can be magnetic or non-magnetic depending on the competing interactions.

\subsection{Spin density wave}
\begin{figure}
\centerline{\epsfig{file=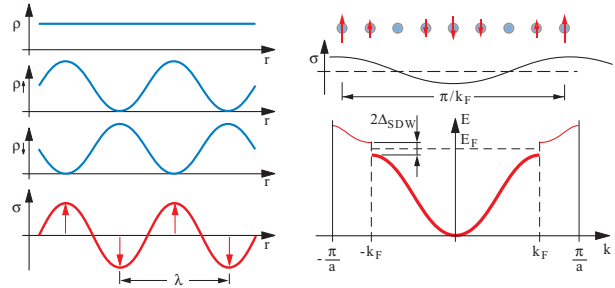,width=8cm}}
\caption{\label{fig:sdw4}
\small Keeping the total charge density $\rho$ constant, the spins up and spins down exhibit a sinusoidal variation: $\rho(r)=\rho_{\uparrow}+\rho_{\downarrow}$. Thus the spins form a density wave with period $\lambda=\pi/k_F$; the Brillouin zone is reduced to $k_F$, leading to a gap $2\Delta_{\rm SDW}$ in conduction band at the Fermi energy $E_F$.}
\end{figure}

\begin{figure}
\centerline{\epsfig{file=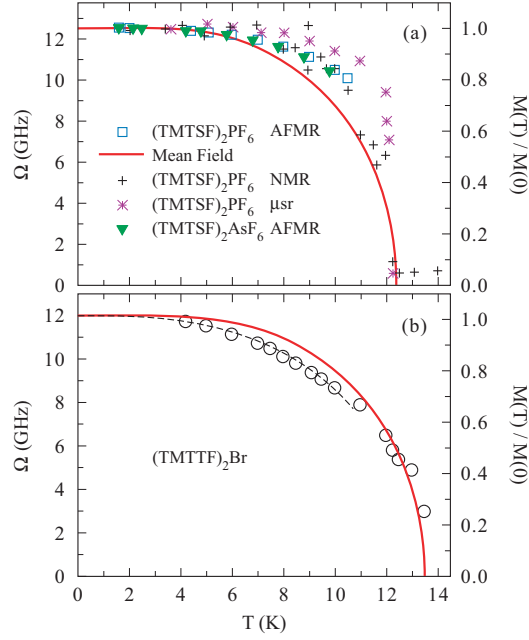,width=7cm}}
\caption{\label{fig:sdw2}
\small (a) Temperature dependence of the low-frequency
zero-field mode $\Omega_{-}$ of (TMTSF)$_2$PF$_6$ determined from measurements of the antiferromagnetic resonance.
The results are compared with the predictions of the mean-field theory,
NMR-measurements (Takahashi et al. 1986),
$\mu$SR-measurements (Le et al. 1993) and
the temperature dependence of $\Omega_{-}$/$\Omega_{-}(0)$ of
(TMTSF)$_2$AsF$_6$.
(b) Temperature dependence of $\Omega_{-}$ of (TMTTF)$_2$Br.
The solid line represents the temperature dependence
of the sublattice magnetization expected in mean-field
theory, the dashed line corresponds to $M(T)/M(0) = 1 - cT^{3}$
with $c = 2.7\times 10^{-4}$~K$^{-3}$ (after Dumm et al. 2000b).}
\end{figure}

\begin{figure}
\centerline{\epsfig{file=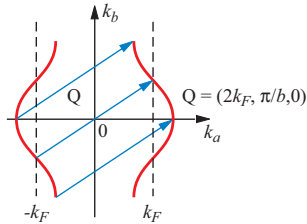,width=4cm}}
\caption{\label{fig:sdw1}
\small Schematic Fermi surface nesting of a
quasi one-dimensional system with interchain coupling in
$b$-direction. For strictly one-dimensional conductors the Fermi surface consists of two planes (dashed lines) separated by $Q=2k_F$. For the quasi one-dimensional example the nesting vector is $Q=(2k_F,\pi/b,0)$.}
\end{figure}

Similar to a CDW, the density of spins up $\rho_{\uparrow}$ and spins down $\rho_{\downarrow}$ can be modulated without affecting the total density of electronic charge $\rho({\bf r})$ as a sketched in Fig~\ref{fig:sdw4}; the spin density wave (SDW) is an antiferromagnetic ground state. While in common antiferromagnets the electronic spins are localized on ions, here the conduction electrons carry the magnetic moment. The magnetic order develops with decreasing temperature according to mean-field theory. This was probed for several quasi-one-dimensional charge-transfer salts by nuclear magnetic resonance (NMR) (Takahashi et al. 1986), muon spin resonance ($\mu$SR) (Le et al. 1993) and measurements of the antiferromagnetic resonance (AFMR) (Torrance et al. 1982, Parkin et al. 1992, Dumm et al. 2000b). In Fig.~\ref{fig:sdw2}a the temperature dependence of the magnetization is plotted for different Bechgaard salts.
The driving force for the SDW formation is the Fermi surface instability with a nesting vector ${\bf Q}=2{\bf k}_F$ as diplayed in Fig.~\ref{fig:sdw1}.
The spatial modulation of the electron spin density leads to a superstructure with period $\lambda=\pi/k_F$, and an energy gap $2\Delta_{\rm SDW}(T)$ opens at the Fermi energy; the gap value increases with decreasing temperature the same way as the magnetization. A close inspection of Fig.~\ref{fig:sdw2}a, however, reveals that the transition is not simply following mean-field behavior, but that the order parameter rapidly increases right below $T_{\rm SDW}$, a tendency known from first-order phase transitions.

\begin{figure}
\centerline{\epsfig{file=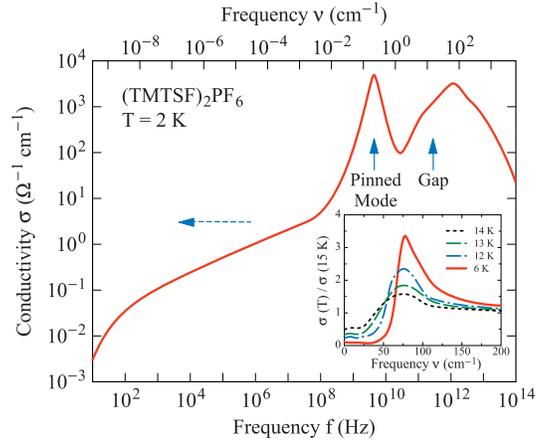,width=7cm}}
\caption{\label{fig:sdw3}
\small Frequency dependent conductivity of (TMTSF)$_2$PF$_6$ in the spin-density-wave phase ($T=2$~K) measured along the stacking direction $a$. The
single-particle gap and the collective mode due to the pinning of the SDW
at imperfections are denoted by the solid arrows. The dashed arrow indicates the range of internal deformations which lead to a broad relaxational behavior. The opening of the SDW gap is demonstrated in the inset where the normalized conductivity for the perpendicular direction is plotted for different temperatures as indicated.}
\end{figure}

The electrical resistivity exhibits a semiconducting behavior below
$T_{\rm SDW}$ in full analogy to the CDW state; the example of \pf6\ is
shown in curve 3  of Fig.~\ref{fig:dc}. If the wavelength of the
density wave modulation $\lambda$ is a multiple of the lattice period,
the density wave of the electronic system is rigidly connected to the
ions. If it is incommensurate with the underlying lattice, it can in
principle freely move. The pinning of the entire SDW on impurities in
the crystal leads to collective transport only above a certain
threshold field, and in an alternating electric field with a resonance
frequency in the GHz range (Gr\"uner 1994). A typical example of a
one-dimensional metal which undergoes an incommensurate  SDW transition
at $T_{\rm SDW}=12$~K is the Bechgaard salt (TMTSF)$_2$PF$_6$. The
electrodynamic response plotted in Fig.~\ref{fig:sdw3} clearly exhibits
the opening of a well-defined gap around $2\Delta_{\rm SDW}=70$~\cm\ as
the temperature is lowered (Degiorgi et al. 1996). The contribution of the collective mode
peaks around the pinning frequency $\nu_0=0.1$~\cm. The spectral weight
of this mode, however, is too small to compensate for the reduction
upon entering the SDW phase; by now the reason for the missing spectral
weight is not completely understood. Albeit the density wave is pinned
to randomly positioned impurities, local deformations lead to an
internal polarization of the mode which results in low-lying
excitations. Such effects are described by a broad relaxation process
in the kHz and MHz range of frequency (Donovan et al. 1994).

While theory explores perfectly one-dimensional systems, in real materials the conducting chains will always be weakly coupled; hence quasi-one-dimensional system are  only a more or less good approximation of the limiting case. As demonstrated in Fig.~\ref{fig:sdw1} partial nesting is still possible and parts or the entire Fermi surface can become gapped below $T_{\rm SDW}$.
The collective response is restricted to the nesting vector which in the quasi-one-dimensional case is not parallel to chain direction any more, but has some perpendicular component. Recent microwave measurement on \pf6\ provide strong evidence that the pinned mode resonance is also present in the $b$-direction (Petukhov and Dressel 2005).

\subsection{Antiferromagnetic spin chain}
\begin{figure}
\centerline{\epsfig{file=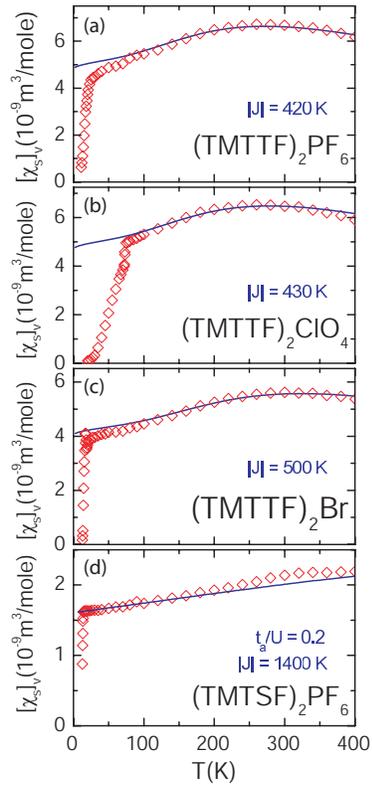,width=4.8cm}}
\caption{\label{fig:chi}
\small Temperature dependence of the spin
susceptibility $(\chi_{\rm s})_{\rm v}$ at constant volume of different
TMT$C$F-salts as obtained by ESR intensity. The lines in (a) to (c) correspond to a
$S = 1/2$ AFM Heisenberg chain with $J= 420$~K, $J= 430$~K and $J= 500$~K, respectively; the line in the lowest frame~(d) corresponds to a fit by an advanced  model with $t_a/U = 0.2$ (Dumm et al. 2000a).}
\end{figure}

One-dimensional spin chains have the tendency to order
antiferromagnetically. This is nicely seen in the spin susceptibility
of Fabre and Bechgaard salts (TMT$C$F)$_2X$ as displayed in
Fig.~\ref{fig:chi}: at high temperatures $\chi(T)$ corresponds to a
spin 1/2 antiferromagnetic Heisenberg chain with exchange constants $J
= 420 - 500$~K. The magnetic coupling $J$ gives the energy scale of the ordering; for significantly lower temperatures, the susceptibility decreases because the the spins
cannot follow the magnetic field any more.

\begin{figure}
\centerline{\epsfig{file=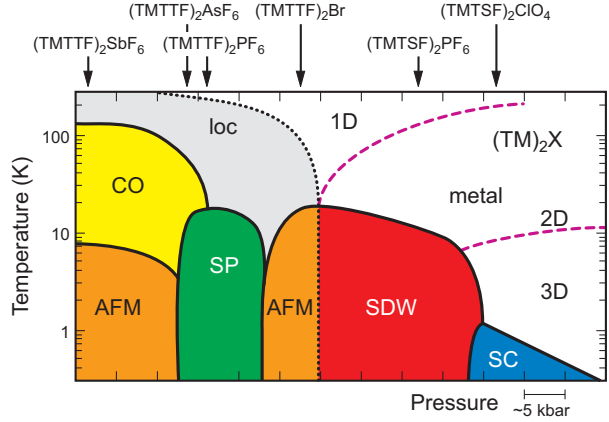,width=8cm}}
\caption{\label{fig:phasediagram}
\small The phase diagram
of the quasi one-dimensional TMTTF and TMTSF salts,
first suggested by J{\'e}rome and coworkers
(J{\'e}rome 1991) and further developed over the years. For the different compounds the ambient-pressure position in the phase
diagram is indicated. Going from the left to the right, the materials get less one-dimensional due to the
increasing interaction in the second and third direction. Here \textsf{loc} stands for charge localization,
\textsf{CO} for charge ordering, \textsf{SP} for
spin-Peierls, \textsf{AFM} for antiferromagnet, \textsf{SDW} for spin density wave, and \textsf{SC} for superconductor. The
description of the metallic state changes from a one-dimensional Luttinger liquid to a two and
three-dimensional Fermi liquid. While some of the boundaries are clear phase transitions,
the ones indicated by dashed lines are better characterized as a crossover. The position in the phase diagram can be tuned by
external or chemical pressure.}
\end{figure}

\begin{figure}
\centerline{\epsfig{file=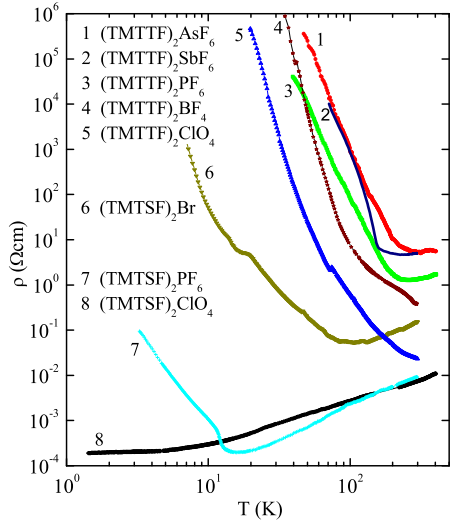,width=6cm}}
\caption{\label{fig:dc}
\small Temperature dependence of the dc resistivity of several Fabre and Bechgaard salts. As the temperature is reduced, the charges become increasingly localized in (TMTSF)$_2$AsF$_6$ and (TMTSF)$_2$PF$_6$,
before the charge-ordered state is entered below 100~K. (TMTSF)$_2$SbF$_6$ shows a transition from a metal-like state directly into the charge-odered state at $T_{\rm CO}=150$~K. (TMTSF)$_2$PF$_6$ undergoes a SDW transition at $T_{\rm SDW}=12$~K. Only (TMTSF)$_2$ClO$_4$ remains metallic all the way down to approximately $T_c=1.2$~K where it becomes superconducting (after Dressel et al. 2001).}
\end{figure}

In all cases, however, a transition to an ordered ground state is
observed by a drop of the spin susceptibility at low temperatures,
which has quite different nature; as summarized in the phase diagram
Fig.~\ref{fig:phasediagram}. For instance, \pf6\ develops a spin
density wave, as just discussed, where the internal magnetic field shifts and significantly broadens the resonance line. Below $T_{\rm SDW}$ all the charge carriers (and thus all the spins) enter a collective state in which the spins form pairs (Fig.~\ref{fig:chi}d). The
tetrahedral anions in (TMTTF)$_2$ClO$_4$ are subject a an anion
ordering at $T_{\rm AO}=72.5$~K which can be identified by a kink in
the temperature-dependent resistivity (curve 5 in Fig.~\ref{fig:dc}).
This lattice rearrangement results in an alternating coupling $J_1$ and
$J_2$ and thus in a singlet ground state. For other non-centrosymmetric anions like ReO$_4$, BF$_4$ and SCN of (TMTTF)$_2X$ similar anion ordering transitions are identified in the range $40~{\rm K}<T_{\rm AO} < 160$~K. Also in (TMTTF)$_2$PF$_6$ the
ground state is non-magnetic, but the reason for this is a spin-Peierls
transition at around 19~K; this will be discussed in more detail in the
following section.

The electrons in (TMTTF)$_2$Br are by far more localized compared to
the one-dimensional metals (TMTSF)$_2X$ as can be seen from
Fig.~\ref{fig:dc}. The antiferromagnetic phase transition at $T_{\rm
N}=13.3$~K is induced by three-dimensional ordering of the
one-dimensional chains of localized spins. ESR experiments presented in
Fig.~\ref{fig:sdw2}b evidence that the magnetization perfectly traces
the mean-field behavior (Dumm et al. 2000b). As depicted in
Fig.~\ref{fig:phasediagram}, the completely insulating
(TMTTF)$_2$SbF$_6$ exhibits a ground state very similar to
(TMTTF)$_2$Br; however the antiferromagnetic phase which develops below
$T_N=8$~K arises out of a charge-ordered state (Yu et al. 2004). With
large enough on-site and nearest-neighbor Coulomb repulsion $U$ and $V$ in
the 1/4-filled system, a charge pattern of alternating  rich and poor
sites ({--} + $\vert$ {--} + $\vert$ {--} + $\vert$ {--} +) is produced at $T_{\rm CO}=156$~K
and hence the ground state is antiferromagnetic. However, taking
electron-lattice coupling into account other charge configurations
become possible (Mazumdar et al. 2000). In particular the ferroelectric
response observed by measurements of dielectric permittivity requires a
coupling term between the electrons in the stack and the charged
counterions (Monceau et al. 2001). The phase diagram
(Fig.~\ref{fig:phasediagram}) infers a competition between the CO- and
SP-order parameter. Increasing pressure leads to a frustration of the
CO resulting from a modified coupling to the counterions, and once it
is sufficiently suppressed, the ground state is singlet (spin-Peierls)
rather than antiferromagnetic (Yu et al. 2004). It remains an open
question, to which extend the second antiferromagnetic state (observed
in (TMTTF)$_2$Br, for instance) is a reentrance of this phase, or
whether it is a distinctly different symmetry breaking.

\subsection{Spin Peierls transition}
While the spin density wave in first approximation does not couple to the lattice, the spin-Peierls (SP) transition is
a magneto-elastic phase transition. Quasi-one-dimensional $S=1/2$ antiferromagnetic spin chains can gain magnetic energy by forming a singlet ($S=0$) ground state. As a result, at low temperature the spin chains dimerize in the  spin-Peierls state (tetramerization of the lattice) as depicted in Fig.~\ref{fig:coso}g. The formation of $S = 0$ spin pairs, which are well localized, yields a non-magnetic
ground state; the spin susceptibility
decreases exponentially below the temperature $T_{\rm SP}$.

\begin{figure}
\centerline{\epsfig{file=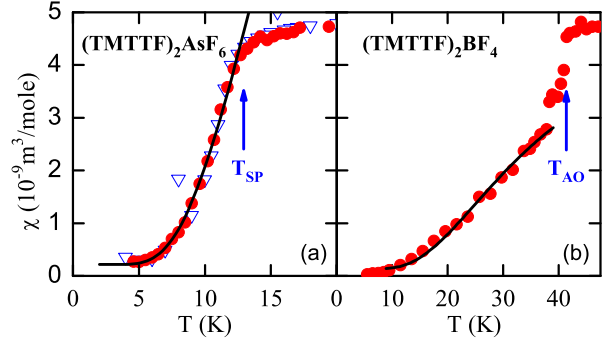,width=8cm}}
\caption{\label{fig:sp1} \small Low-temperature behavior of the spin susceptibility of (TMTTF)$_2$AsF$_6$ and (TMTTF)$_2$BF$_4$ obtained from ESR (solid circles) and SQUID measurements (opend triangles). (a)~At $T_{\rm SP}=13$~K the susceptibility of (TMTTF)$_2$AsF$_6$ decreases exponentially indicating the spin-Peierls transition to a non-magnetic ground state. The solid line corresponds to a fit by a mean-field model. (b)~(TMTTF)$_2$BF$_4$ undergoes a first-order phase transition at $T_{\rm AO}=41$~K due to the ordering of the tetrahedral anions that results in a rapid drop of $\chi(T)$. At low temperatures the susceptibility in the spin ordered state also follows an activated behavior.}
\end{figure}

In Fig.~\ref{fig:sp1}a the example of (TMTTF)$_2$AsF$_6$ is presented
which enters the SP state at $T_{\rm SP}=13.1$K. The fit of the
experimental data by a mean-field theory of Bulaevskii (Bulaevskii
1969) yields a ratio of $\gamma=J_2/J_1$ between the inter- and
intra-spin-dimer couplings with $\vert J_1\vert=423$~K. The
singlet-triplet gap is estimated to be $\Delta_{\sigma}(0)=22$~K in
good agreement with mean-field prediction: $2\Delta_{\sigma}(0)/T_{\rm
SP}=3.53$. It is obvious from Figs.~\ref{fig:sp1}a that well above the
actual transition temperature, the spin susceptibility is already
reduced to one-dimensional lattice fluctuations (Dumoulin et al. 1996, Bourbonnais and Dumoulin 1996). This is even more
obvious in (TMTTF)$_2$PF$_6$ where fluctuations are evident almost up
to 100~K as becomes obvious from Fig.~\ref{fig:chi}a (Dumm et al.
2000a).

Already the investigation of the vibrational mode $\nu_3({\rm A}_g)$
summarized in Fig.~\ref{fig:co3}d gave evidence for the spin Peierls
transition in (TMTTF)$_2$AsF$_6$. Below $T_{\rm SP}=13$~K the mode
splitting decreases which indicates a reduction of the charge
disproportionation. Most important, this observation evidences the
coexistence of both ordering phenomena; implying that in addition to
the spin arrangement depicted in Fig.~\ref{fig:coso}g the charge
remains modulated in a ({--} + $\vert$ {--} + $\vert$ {--} + $\vert$ {--} +)
pattern.

\subsection{Spin order}
In contrast to a spin-Peierls transition, where spin-phonon interaction
is responsible for the ordering, the tetramerization can occur due to
structural changes, like the ordering of tetrahedral anions ClO$_4$,
ReO$_4$ or BF$_4$. The result is a charge arrangement
({--} + $\vert$ + {--} $\vert$ {--} + $\vert$ + {--}) as plotted in
Fig.~\ref{fig:coso}f. The alternating exchange constants $J_1$ and $J_2$ lead to a
spin ordering with a singlet ($S=0$) ground state. The first-order
phase transition is accompanied by a step-like decrease in the spin
susceptibility. The further decrease in $\chi(T)$ can be well described
by Bulaevskii's model (Bulaevskii 1969) of an alternating spin chain
with a ratio of the exchange constants $\gamma=0.9$ and 0.8 for
(TMTTF)$_2$BF$_4$ and (TMTTF)$_2$ClO$_4$, respectively; the
singlet-triplet gaps are $\Delta_{\sigma}=52$~K and 84.5~K . The sudden
decrease of the electrical resistivity, as demonstrated by the kink in
curves 4 and 5 of Fig.~\ref{fig:dc} for the examples of
(TMTTF)$_2$BF$_4$ and (TMTTF)$_2$ClO$_4$, indicates reduced  scattering in the ordered state and rules out a change in the density
of states due to the formation of a pseudogap.

Interestingly, while in general for non-centosymmetric counter ions in (TMTTF)$_2X$ anion ordering is observed, only for $X$ = ClO$_4$, ReO$_4$ and BF$_4$ a spin gap $\Delta_{\sigma}$ opens in the non-magnetic anion-ordered ground state, for (TMTTF)$_2$SCN  the anion order is accompanied by a charge order, but not by a spin order.

\section{Outlook}
No doubt, one-dimensional physics matured from a toy model to an
extremely active field of theoretical and experimental research,
spanning a broad range from quantum gases to condensed-matter physics
and semiconductor technology. A large variety of novel and exciting phenomena can
be investigated in these systems. In one-dimensional metals collective
modes replace the single-particle excitations common to
three-dimensional conductors and successfully described by Landau's Fermi liquid
concept of interacting electrons. Another property typical for
low-dimensional solids is their susceptibility to symmetry breaking
with respect to the lattice, the charge and the spin degree of freedom.
Broken-symmetry ground states imply that the system becomes stiff,
because the modulation of the order parameter costs energy; therefore
collective modes appear at low energies. In the case of magnets, the
loss of rotational invariance leads to a magnetic stiffness and spin
waves. In superconductors the gauge symmetry is broken, but due to the
Higgs mechanism the Goldstone mode is absent at low frequencies and
shifted well above the plasma frequency. In the examples above, we were
dealing with translational invariance which is lowered in crystals due to
charge ordering phenomena.

Charge density waves drive a metal to an insulator for the Fermi
surface becomes instable; the pinned-mode resonance can nicely be
detected in the GHz using a variety of high-frequency and optical
techniques. Purely electronic correlations between adjacent sites can
cause charge disproportionation. Organic conductors are suitable
realizations to investigate the properties at the metal-insulator
transitions. The neutral-ionic transition observed in mixed-stack
one-dimensional organic charge-transfer salts can be a pure change of
ionizity, but commonly goes hand in hand with a Peierls distortion.
This can be seen in a softening of the low-frequency phonon modes above
the phase transition.

In general, the magnetic coupling is weaker compared to electronic effects, hence the ordering occurs
at lower temperatures. The competition of electronic, magnetic and
phonon interaction is responsible that a particular ground state
develops, which can range from a nonmagnetic singlet, spin-Peierls, an
antiferromagnet to a spin-density wave state, as depicted in the phase
diagram of the Bechgaard and Fabre salts (Fig.~\ref{fig:phasediagram}).
The well-balanced interplay of these interactions calls for further exploration and will remain an active field of experimental and theoretical research for the years to come.
Unfortunately, these organic solids permit neutron scattering
experiments only in a very limited way. Thus the exploration of
spin-wave excitations is still an open issue.

\section*{Acknowledgements}
During the last years, we enjoyed collaborations and discussions with
S. Brown, L. Degiorgi, N. Drichko, M. Dumm, A. Girlando, G. Gr\"uner and S. Tomi\'c. We thank N. Drichko, M. Dumm and S. Yasin for providing unpublished data.

\section*{References}
\setlength{\itemsep}{0mm}
\setlength{\parskip}{1.2mm}
{\small
\begin{trivlist}
\item
Basista H, Bonn DA, Timusk T, Voit J, J\'erome D, Bechgaard K (1990)
Far-infrared optical properties of tetrathiofulvalene-tetracyanoquinodimethane (TTF-TCNQ). Phys Rev B 42: 4008-4099
\item
Batail P (ed) (2004)
Molecular Conductors. Thematic Issue of Chemical Reviews 104: 4887-5781
\item
Bourbonnais C, Dumoulin B (1996)
Theory of lattice and electronic fluctuations in weakly localized spin-Peierls systems. J Phys I (France) 6: 1727-1744
\item
Bulaevskii LN (1969) Magnetic susceptibility of a chain of spins with
antiferromagnetic interaction. Sov Phys Solid State 11: 921-924
\item
Chow DS, Zamborszky F, Alavi B, Tantillo DJ, Baur  A, Merlic CA, Brown SE (2000)
Charge ordering in the TMTTF family of molecular conductors.
Phys Rev Lett 85: 1698-1701
\item
Cohen MJ, Coleman LB, Garito AF, Heeger AJ (1974)
Electrical conductivity of tetrathiofulvalinium tetracyanoquinodimethan (TTF)(TCNQ). Phys Rev B 10: 1298-1307
\item
Coleman LB, Cohen  MJ, Sandman DJ, Yamagishi FG, Garito AF, Heeger  AJ (1973)
Superconducting fluctuations and the Peierls instability in an organic solid. Solid State Commun 12: 1125-1132
\item
Davies JH (1998)
The physics of low-dimensional semiconductors.
Cambridge University Press, Cambridge
\item
Degiorgi L, Dressel M, Schwartz A, Alavi B, Gr\"uner G (1996)
Direct observation of the spin-density-wave gap in (TMTSF)$_2$PF$_6$.
Phys Rev Let 76: 3838-3841
\item
Denoyer F, Com\`es F, Garito AF, Heeger AJ (1975)
X-ray-diffuse-scattering evidence for a phase transition in tetrathiafulvalene tetracyanoquinodimethane (TTF-TCNQ).
Phys Rev Lett 35: 445-449
\item
Donovan S,  Kim Y, Degiorgi L, Dressel M, Gr\"{u}ner G,  Wonneberger W (1994)
The electrodynamics of the spin density wave ground state: optical experiments on (TMTSF)$_2$\-PF$_6$. Phys Rev B 49: 3363-3377
\item
Dumoulin B, Bourbonnais C, Ravy S, Pouget JP, Coulon C (1996)
Fluctuation effects in low-dimensional spin-Peierls systems: theory and experiment. Phys Rev Lett 76: 1360-1363
\item
Dressel M, Kirchner S, Hesse P, Untereiner G,  Dumm M, Hemberger J,
Loidl A,  Montgomery L (2001) Spin and charge dynamics in Bechgaard
salts. Synth Met 120: 719-720
\item
Dressel M, Gr\"uner G (2002)
Electrodynamics of Solids.
Cambridge University Press
\item
Dressel M (2003)
Spin-charge separation in quasi one-dimensional organic conductors.
Naturwissenschaften 90: 337-344
\item
Dumm M,  Loidl  A,  Fravel BW,  Starkey KP,  Montgomery L,  Dressel M  (2000a)
Electron-spin-resonance studies on the organic linear chain compounds (TMT$C$F)$_{2}X$ ($C$=S, Se and $X$=PF$_6$, AsF$_6$, ClO$_4$, Br).
Phys Rev B 61:  511-520
\item
Dumm M, Loidl A, Alavi B, Starkey KP,  Montgomery L, Dressel M (2000b)
Comprehensive ESR-study of the antiferromagnetic ground states in
the one-dimensional spin systems (TMTSF)$_{2}$PF$_6$, (TMTSF)$_{2}$AsF$_6$, and (TMTTF)$_{2}$Br.
Phys Rev B 62: 6512-6520
\item
Dumm M, Abaker M, Dressel M (2005)
Mid-infrared response of charge-ordered quasi-1D organic conductors (TMTTF)$_2X$. J Phys IV (France)  131: 55-58
\item
Farges JP (ed.) (1994)
Organic conductors.
Marcel Dekker, New York
\item
Giamarchi T (2004)
Quantum physics in one dimension.
Oxford University Press, Oxford
\item
Goldstone J (1961)
Field theories with `superconductor' solution. Nuovo cimento 19: 154-164
\item
Gr\"uner G (1994) Density waves in solids, Addison-Wesley, Reading, MA
\item
Heeger AJ, Garito AF (1975) The electronic properties of TTF-TCNQ.
in:  Keller HJ (ed) Low dimensional cooperative phenomena, Plenum, New York, 89-123
\item
Himpsel FJ, Kirakosian  A, Crain JN, Lin JL, Petrovykh DY (2001)
Self-assembly of one-dimensional nanostructures at silicon surfaces.
Solid State Commun 117: 149-157
\item
Horiuchi S, Okimoto Y, Kumai R, Tokura  Y (2000)
Anomalous valence fluctuation near a ferroelectric transition in an organic charge-transfer complex. J Phys Soc Jpn 69: 1302-1305
\item
Horiuchi S, Okimoto Y, Kumai R, Tokura  Y (2003)
Quantum phase transition in organic charge-transfer complexes. Science 299: 229-232
\item
Ishiguro T, Yamaji K, Saito G (1998)
Organic superconductors. 2nd edition,
Springer, Berlin
\item
J{\'e}rome D, Mazaud A, Ribault M,  Bechgaard K (1980)
Superconductivity in a synthetic organic conductor (TMTSF)$_2$PF$_6$.
J.\ Physique Lett.\  41: L95-97
\item
J{\'e}rome D,   Schulz HJ (1982)
Organic conductors and superconductors.
Adv Phys 31: 299-490
\item
J{\'e}rome D (1991)
The Physics of Organic Conductors.
{Science} 252:  1509-1514
\item
Kagoshima S, Nagasawa  H, Sambongi T (1988) One-dimensional conductors Springer, Berlin
\item
Koshihara SY, Takahashi Y, Saki H, Tokura Y, Luty T (1999) Photoinduced cooperative charge transfer in low-dimensional organic crystals. J Phys Chem B 103: 2592-2600
\item
Le LP, et al. (1993)
Muon-spin-rotation and relaxation studies in (TMTSF)$_2X$ compounds.
Phys Rev. B 48: 7284 - 7296
\item
Lieb EH, Mattis DC (ed.) (1966)
Mathematical physics in one dimension.
Academic Press, New York
\item
Maekawa S, Tohyama T, Barnes SE, Ishihara S, Koshibae W, Khaliullin G (2004)
The physics of transition metal oxides.
Springer-Verlag, Berlin
\item
Masino M, Girlando A, Brillante A, Della Valle RG, Venuti E, Drichko N, Dressel M (2006) Lattice dynamics of TTF-CA across the neutral ionic transition. Chem Phys 325: 71-77
\item
Mazumdar S, Clay RT, Cambell DK (2000) Bond-order and charge-density
waves in the isotropic interacting two-dimensional quarter-filled band
and the insulating state proximate to organic superconductivity. Phys
Rev B 62: 13400-13425
\item
Monceau P (Ed.) (1985)
Electronic properties of inorganic quasi-one-dimensional compounds, Part I/II.
Reidel, Dordrecht
\item
Monceau P, Nad FY, Brazovskii S (2001)
Ferroelectric Mot-Hubbard phase of organic (TMTTF)$_2X$ conductors.
Phys Rev Lett 86: 4080-4083
\item
Moritz H, St\"oferle T, K\"ohl M, Esslinger T (2003)
Exciting collective oscillations in a trapped 1D gas.
Phys Rev Lett  91: 250402-1-9
\item
Nad F, Monceau P, Fabre J (1999)
High dielectric permittivity in quasi-one-dimensional organic compounds (TMTTF)2X: Possible evidence for charge induced correlated state.
J Phys (Paris) IV 9: Pr10-361-364
\item
O'Connell M (2006)
Carbon nanotubes.
Taylor \&\ Francis, Boca Raton
\item
Overhauser AW (1962)
Spin density waves in an electron gas.
Phys Rev 128: 1437-1452
\item
Parkin SSP, Scott  JC, Torrance JB, Engler EM (1982)
Antiferromagnetic resonance in tetramethyltetrathiafulvalene bromide [(TMTTF)$_2$Br].
Phys Rev B 26: 6319-6321
\item
Petukhov K, Dressel M (2005)
Collective spin-density-wave response perpendicular to the
chains of the quasi one-dimensional conductor (TMTSF)$_2$PF$_6$.
Phys Rev B 71: 073101-1-3
\item
Seo H, Fukuyama H (1997)
Antiferromagnetic phases of one-dimensional quarter-filled organic conductors.
J Phys Soc Jpn 66: 1249-1252
\item
Takahashi T, Maniwa Y, Kawamura H, Saito G (1986)
Determination of SDW characteristics in (TMTSF)$_2$PF$_6$ by $^1$H-NMR analysis. J Phys Soc Jpn 55: 1364-1373
\item
Torrance JB, Vazquez JE, Mayerle  JJ,  Lee  VY (1981)
Discovery of a neutral-to-ionic phase transition in organic materials. Phys Rev Lett  46: 253-257
\item
Torrance JB, Pedersen HJ, Bechgaard K (1982)
Observation of antiferromagnetic resonance in an organic superconductor
Phys Rev Lett  49: 881-884
\item
Vuleti\'c T,  Korin-Hamzi\'c B, Ivek T, Tomi\'c S, Gorshunov B,
Dressel M, Akimitsu J (2006)
The spin-ladder and spin-chain system
(La,Y,Sr,Ca)$_{14}$Cu$_{24}$O$_{41}$:
electronic phases, charge and spin dynamics.
Phys Rep 428: 169-258
\item
Yu W, Zhang F, Zamborszky F, Alavi B, Baur A, Merlic CA, Brown SE
(2004) Electron-lattice coupling and broken symmetries of the molecular
salt (TMTTF)$_2$SbF$_6$. Phys Rev B 70: 121101-1-4
\item
Zeller HR (1973)
Electronic Properties of one-dimensional solid state systems.
in: Queisser HJ (ed) Festk\"orperprobleme (Advances in Solid State Physics) Vol 13, Pergamon, New York, p. 31
\item
Zeller HR (1975)
Electrical Transport and spectroscopical studies of the Peierls transition in
K$_2$[Pt(CN)$_4$]Br$_{0.30}\cdot$3H$_2$O.
in: Keller HJ (ed) Low dimensional cooperative phenomena, Plenum, New York, 215-233
\end{trivlist}
}
\renewcommand{\baselinestretch}{1}\normalsize

\newpage

\end{document}